# Spectrophotometers for Labs: a Cost-efficient Solution based on Smartphones

Carlos Balado Sánchez[1], Rebeca P. Díaz Redondo[1],
Ana Fernández Vilas[1] and Angel M. Sánchez Bermúdez[2]

[1]Information & Computing Lab., AtlantTIC Research Center
School of Telecommunications Engineering. University of Vigo, 36310, Vigo (Spain)
{iclab, rebeca, avilas}@det.uvigo.es

[2]Department of Chemical Engineering. University of Vigo, 36310, Vigo (Spain)
asanchez@uvigo.es

**Abstract**

In this paper we introduce a proposal to provide students in labs with an alternative to the traditional visible range spectrophotometers, whose acquisition and maintenance entails high costs, based on smartphones. Our solution faced two aspects. On the one hand, the software for the smartphone, able to perform the typical functionalities of the traditional spectrophotometers. On the other hand, the portable peripheral support needed to capture the images to be analyzed in the smartphone. The promising results allow this solution to be applied in Bring Your Own Devices (BYOD) contexts.

**Keywords:** VIS Spectrophotometer, BYOD, scientific equipment for labs, Beer-Lambert law, Apache Cordova,

## 1. Introduction

As it is noted in [1], laboratory activities increase students' achievements and interest in the subject matters, especially in STEM disciplines. At this respect, Web and Cloud learning environments and technologies provide many opportunities to enhance learning experience for both on-campus and off-campus students [2]. Online distance learning has been accompanied by the implementation of virtual and remote labs so that distance students can carry out experiments as on-campus students and even on-campus students and teacher gain flexibility and freedom with 24/7 access to learn at one's own pace [3]. On the other hand, consumerization of IT has been accompanied by the Bring Your Own Device (BYOD) concept, first proposed in Intel, that allows employees to utilize their personally-owned technology, devices to stay connected to, access data from, or complete tasks for their organizations. As a translation form the professional to the education scenarios. BYOD model promises a capillary diffusion of mobiles in schools, thanks to the extreme pervasiveness these devices have reached amongst youngsters. According to that reality, BYOD has been tested at different educative levels [4, 5, 6] and that would be appropriate for both secondary and higher education. For instance, in [6] a BYOD-based experimentation environments is proposed for working on acoustics and noise monitoring that combines mobile sensors, a specific app for noise measurement collection, and a Learning Management System (LMS) for course organization activities.

In this paper, we propose to take advantage of the BYOD approach by proposing an experimentation environment which combines mobile sensor, mobile app and Web-based learning environments in the field of engineering education at University level for chemical analysis and particle physics, more specifically in the Visible range spectrophotometers (VIS spectrophotometers). These devices are usually needed in academic labs. However, its high acquisition and maintenance cost, with prices between 700 EUR (856 USD) and 13.600 EUR (16.633 USD), explain why some institutions cannot afford having this kind of instrumental for each student to practice. What is more, especially if we talk of non-experienced students, the risk of the equipment is higher and so the maintenance costs. As an alternative, students often work in groups, with less time to experiment by themselves and the equipment.

In this paper, we explore the use of smartphones as a viable alternative to this sophisticated and expensive equipment. Our approach is twofold. On the one hand, the software needed to perform the usual functionality of the VIS spectrophotometers. For this purpose, we have opted for solution that combines open software and specific software developed to access the different features of the smartphone. Although the open software was obtained from the online project Spectral Workbench [7], which usually works as a web service, we decided to provide a hybrid solution. Thus, our approach, developed using Apache Cordova IDE [8] and Node.js [9], offers highly versatile possibilities unifying web services and mobile applications to be functional in both web browsers and smartphones of different range. On the other hand, we have also designed a removable peripheral device needed to support the right capture of the images needed to be processed. This device can be easily created using a 3D printer and, what is more important, do not interfere with the normal use of the smartphone. Therefore, our approach is coherent with the recent approaches of BYOD that have been tested at different educative levels and that would be appropriate for both secondary and higher education.

The paper is structured as follows. Section 2 summarizes those technical aspects needed for the proposal, such as the functionality of the VIS spectrophotometer, the Spectral Workbench project, the Apache Cordova IDE and the Noje.js platform. The system is described in Section 3, whereas Section 4 summarizes the results obtained. In Section 5 we discuss the different alternatives to our approach and a brief comparison. Finally, conclusions and further work is described in Section 6.

## 2. Background

### 2.1 Spectrophotometer

A spectrophotometer [10-11] is a tool for chemical analysis that uses the wavelengths to measure the relationship between values of the same photometric magnitude. In other words, it measures the amount of intensity of light absorbed after passing through a sample solution. Thus, it works with the following variables: transmittance or the relationship between intensity of the input and the output's light beam intensity given a specific wave frequency; absorbance or the amount of light intensity absorbed by the sample; and the concentration of the solution. Using these variables, a beam of light can be used to measure the relationship between the intensity of the incident beam on the sample and the intensity of the outgoing beam to obtain absorbance and concentration values in optic spectrophotometric methods. For this to be possible, spectrophotometers need the following elements to work (as Fig. 1 depicts): a constant energy source of radiation; a wavelength selector to choose a specific wavelength to work with, such as filters and monochromators; a detector that transforms the radiation into electric pulses; and a signal processor to work with the results. In fact, they are the mechanical elements needed to recreate this tool with the smartphone.

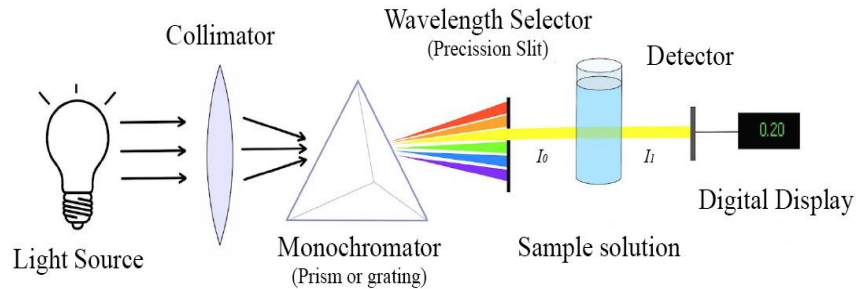

Fig. 1. Diagram of a spectrophotometer internal behavior.

**2.1      Spectral Workbench**

Spectral Workbench [7, 12-14] is an open source web service that functions as a spectrophotometry library around one Public Lab's communities focused on the development of do-it-yourself work techniques. Its main purpose bringing science close to anyone who wants to experience or learn no matter what their science background is.

Spectral Workbench offers several web tools to capture spectra with a webcam as well as features to visualize, manipulate and store spectra from a browser web page. The code used to transform image into data is offered as an open resource written in JavaScript and ready for its use on web browsers. Precisely, this is the software we have taken advantage of: (i) part of the API to transform images into graphics and (ii) to perform some operations and transformation on these graphics. Contrarily to other approaches that also use this open software, our proposal goes further than interacting with Spectral Workbench as a simple database where store and visualize spectrophotometric images.

**2.2 Software Development support**

**Apache Cordova** [8] is a free-license mobile development framework that allows the use of standard web technologies such as HTML5, CSS3 and JavaScript for multiplatform development. The applications are executed within specific wrappers for each platform and use compatible standards to access the device's API and its elements, such as the sensors, the camera [15], the file system or the network status. Precisely, we use Apache Cordova to adapt the server web-based JavaScript code of Spectral Workbench to smartphones but still allowing the access to the different elements of the smartphone, such as the camera and the file system. What is more, because of this software development platform, this implementation was designed combining the power of Web-based technologies with the versatility of mobile applications with the added advantage of the abstraction of the device's native language and, therefore, of the operating system where the code is compiled.

**Node.js** [9], created in 2009 by Ryan Lienhart Dahl, is a cross-platform, open-source runtime environment supported by an event-oriented architecture. It is based on the ISO 16262 standard specification, ECMAScript [16], that uses different technologies such as JavaScript (Netscape) or JScript (Microsoft) to create network programs such as web servers. Its package npm [17] (Node Package Manager) is especially relevant since it allows the adaptation of packages to the application's code. In fact, the Apache Cordova command-line interface is one of the packages offered by npm, as well as most of the Apache Cordova plugins. Finally, it is also possible to share any developed package with other npm users.

## 3. System Description

As previously mentioned, our proposal consist of two parts or modules, as Fig. 2 shows: a physical structure and a software module, both of them specifically created to work with smartphones. The hardware or peripheral module was designed to transforms light from an arbitrary source into spectra of wavelengths of the visible range to allow the camera of the mobile device to feed light data into the application. The software module was designed and implemented to work on smartphones, with the aim of allowing users to perform the most usual operations with spectrophotometers: photograph spectra with the camera; load, visualize and manipulate data from the sample's spectra; and perform calibrations and adjustments necessary to work with these photographs.

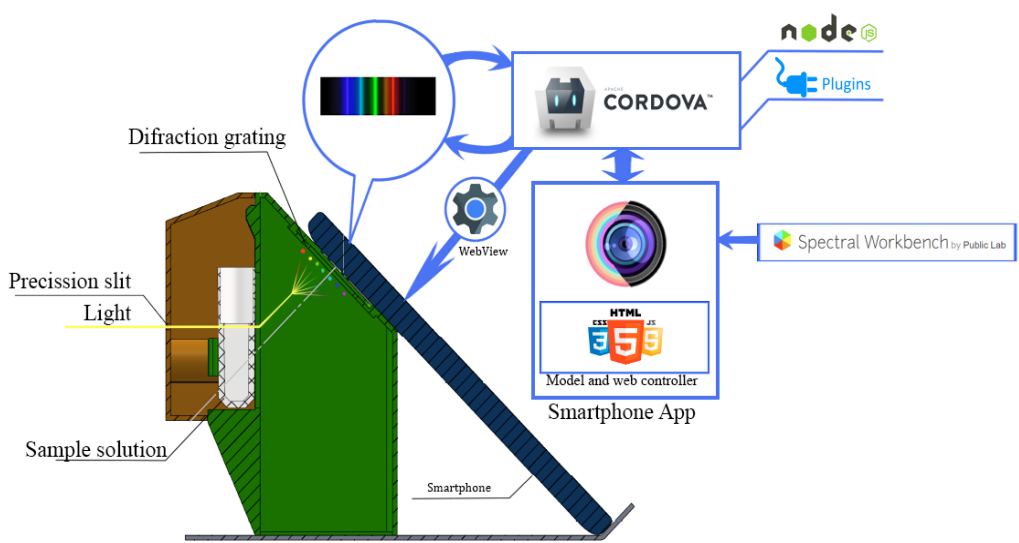

Fig. 2. VIS spectrophotometer scheme

### 3.1 Spectrophotometer software module

The main challenge of the software module is the implementation and adaptation of the tools provided by the Spectral Workbench web service to a smartphone application. These tools, written in JavaScript code, cannot be used by a standard mobile development framework without being rewritten in the respective device's native language (Java, in Android's case) or adding some kind of external plugins that translates or allows the device to understand and compile code of a foreign language.

In order to find a more transparent and suitable solution, we decided to use the device's WebView (internal browser), which supports the programming tools of the web browser. This solution also allows using the smartphone API tools. Consequently, the software implementation will be independent from the smartphone model and, what is more, from the underlying operating system.

When selecting the most suitable development framework, we decided to work with Apache Cordova and Node.js. The former supports JavaScript, HTML5 and CSS in mobile applications. Besides, its command line tools are implemented using the execution environment for

JavaScript Node.js, distributed by its default package manager, npm. This combination unifies the tools of both the web application and the mobile application by using a series of plugins that allow access to the smartphone API. This is important since working with the smartphone API make easy to interact with different elements (camera, storage system, contact list, etc.), which are accessed throught npm. Additionally, and since the most relevant platforms have been added to the Apache Cordova framework, this allows the programmer to withdraw from the model or the operating system of the smartphone and focus on the functionality to develop. Besides, the application can also run in PC browsers although, for obvious reasons, this does not allow the use of the plugins to interact with the smartphone elements.

Therefore, the Apache Cordova Camera plugin provides access to the camera and the image gallery of the smartphone. Once the image of the visible range spectrum is taken and stored, the API of the Spectral Workbench web service obtains intensity values in JSON format of the three color channels (R, G and B) to be graphically represented.

Later, the user can select some of the implemented operations, such as calculating the solute values of concentration, transmittance and absorbance, the visualization of the absorbance versus the concentration graph of single samples or generating a calibration line using an arbitrary number n of solute samples with different concentrations.

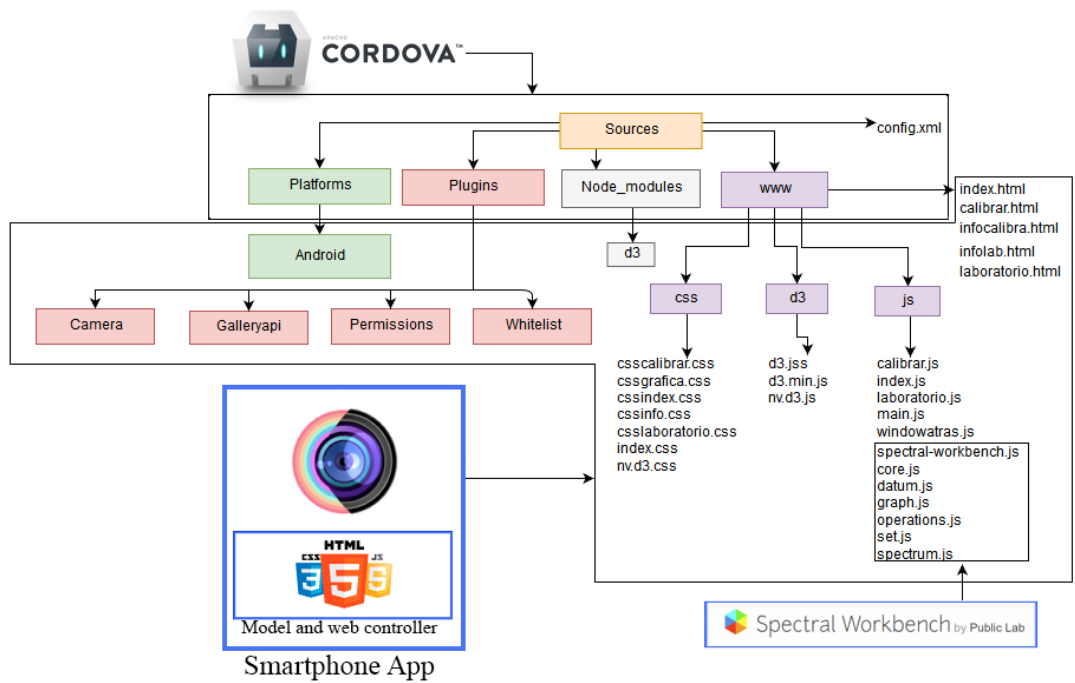

Fig. 3. Application's block diagram.

The common structure of an Apache Cordova project is depicted in Fig. 3, as well as other modules specifically developed for this project. The Sources folder contains the elements needed to create a web-based system and to allow interacting with the features of the device (smartphone): the Platforms subdirectory, where the target system is referenced; the Plugings subdirectory, where all the installed plugins are stored; the WWW subdirectory, where all the logic is programmed and that contains all the JavaScript, HTML5 and CSS files; and, finally, a folder named Node modules, where other necessary files for additional Apache Cordova modules are stored. The information and permissions for the mobile device are managed within the config.xml file, also stored in the Sources root folder.

**3.2 Physical structure of the smartphone**

To obtain the images necessary for the application, we need an external device to diffract and collimate the light beams that pass through the samples. With this aim, we designed a device or physical structure that consists of two moving parts printed with 3D technology that are joined using magnets, as Fig. 4 shows.

The first one, in green in Fig.4, is the support for the smartphone that also holds the sample and attaches a diffraction grating to separate the incident light into a visible spectrum. Both a two-axis diffraction grating and a single-axis one were tested. The latter was chosen since the pattern it generates is much easier to identify with the smartphone camera. The second piece, in yellow in Fig.4, closes the set and has a 0,18nm precision slit that collimates the light beams

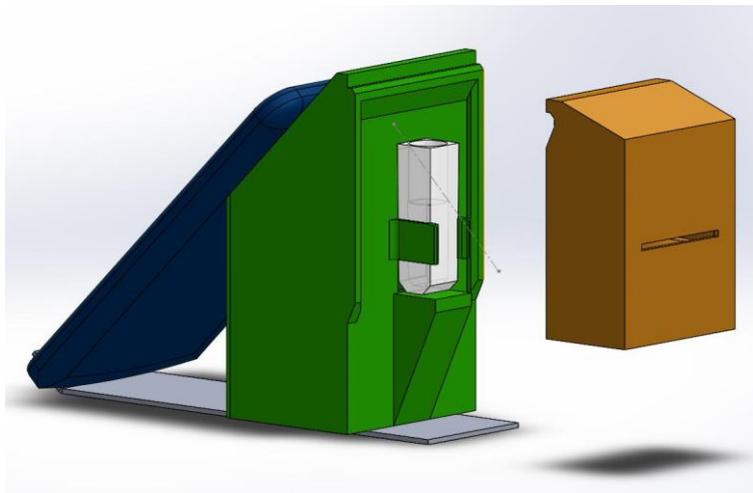

Fig. 4. 3D model of the image obtaining peripheral.

## 4. Results

In order to check if the proposal meets the proposed requirements, different tests were developed. Regarding the usability, several professors that normally use spectrophotometers tested the mobile applications in order to assess the visual design, the developed functionality and the process from taking pictures to performing different operations with them. This feedback was extremely important to improve the software design and the interface.

Additionally, we also perform different experiments using different smartphones to check the portability of the software solution and the performance. Fig. 5 shows the appearance of the software application in two different smartphone models. Of course, and as expected, the highest the processing power of the smartphone, the fastest the application can process the data and display the results.

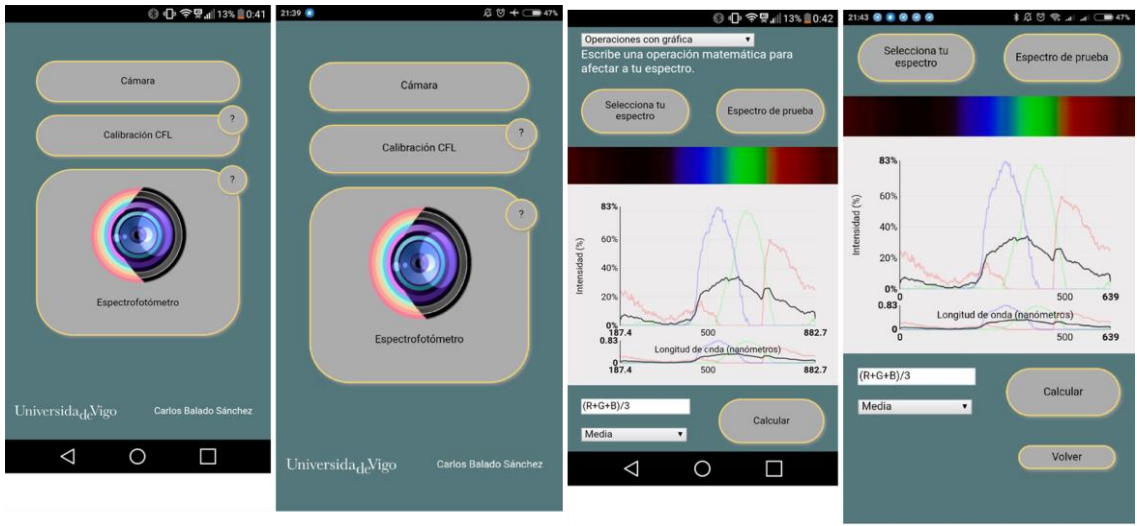

Fig. 5. User interface comparison of the application between a LG-K10 mobile phone and a Xiaomi Redmi Note 4.

Regarding the software module, and after performing different functionality tests on each individual module, the whole application was also tested using the web browsers Chrome and Firefox as debug platforms, thanks to the hybrid nature of the application.

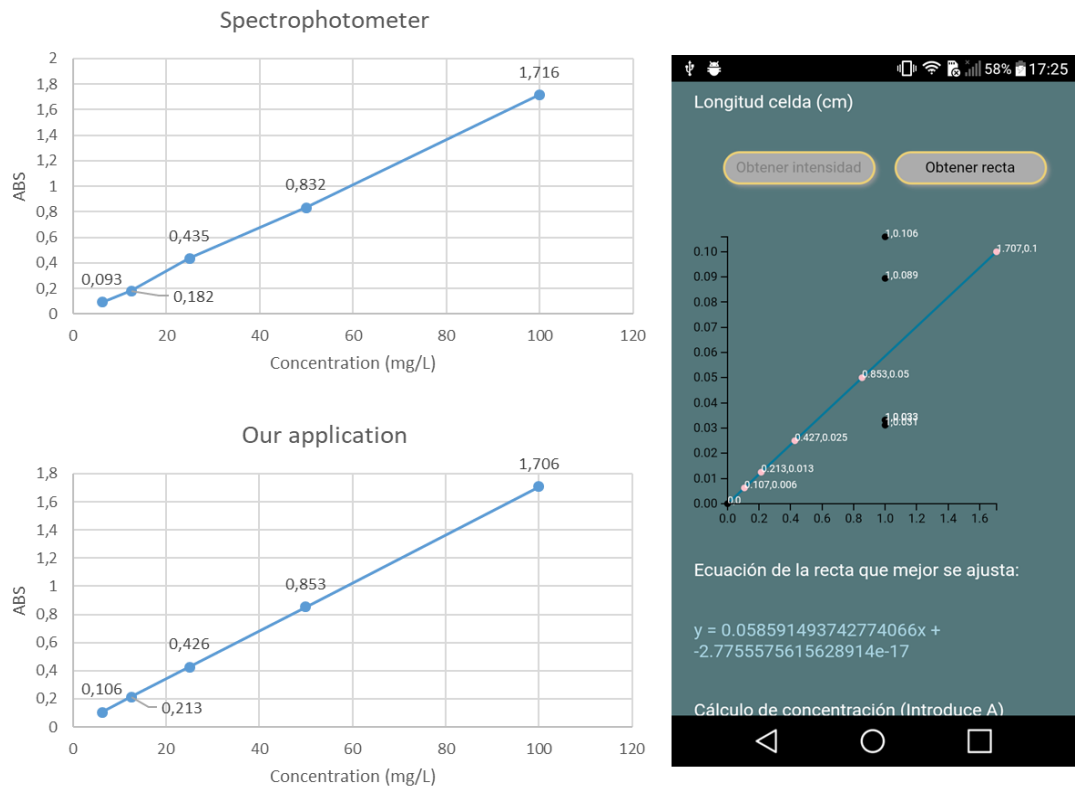

Fig. 6. Comparison between the calibration line with n values obtained with a LG K10 mid range mobile and a laboratory spectrophotometer.

Besides, the Spectral Workbench tool was used to process images and to perform some operations in order to check if the results obtained in our device were the same to the ones obtained with the Spectral Workbench. As result, the quality of the calibration function of the application has been checked to determine if the data is represented correctly and on the other hand it has been checked whether the data itself is shown reliably. In both cases, the degree of error lies between 1.41% and 0.12% after calibration, which seems precise enough considering that an error of one pixel on the device screen is equivalent in most cases to an error rate of 0.71%.

Finally, we also performed tests to compare the values obtained with a Shimadzu UV-1800 spectrophotometer and the smartphone application. With this aim, we made measurements on six samples of food dye solutions of Amaranth red tone (used frequently for this kind of tests in academic environments) and water at concentrations of 100%, 50%, 25%, 12% and 6.25%. The results obtained with both systems (spectrophotometer and our application) are summarized in Fig. 6, as well as the absorbance versus concentration graph (calibration line with n values) obtained with both methods.

The results are clearly promising. With high concentrations (25% and higher) the deviation of the application with respect to the spectrophotometer is between 2.3% and 5.8%. This percentage increases with lower concentrations: 13% and 17% with 12% and 6.25% levels of dye, respectively.

## 5. Related Work and Discussion

There are currently different alternatives when using the mobile device as a spectrophotometric tool. Although there are different applications that serve as databases or try to use the mobile device as a complement to a laboratory spectrophotometer, the most interesting alternatives are those that try to make the smartphone independent of the physical tool by using peripherals and elements external to the device.

However, the main problem when converting a smartphone into a spectrophotometer is that the mobile device does not have the necessary tools to diffract the light and select a wavelength to be able to make the necessary calculations on the samples to be evaluated. Thus, a possible solution is using some material that can diffract the light (for example, the polycarbonate layer of a CD or a DVD) on the camera lens of the mobile device. Other option is designing an artifact able to collides light and avoid interference with natural light. Spectral Workbench proposes a solution, Fig. 7(a), that is adhered and fixed to the smartphone. This kind of approach solves the data acquisition problem, but the cost is too high: some of the characteristics of the smarphone, the camera, the smartphone's cover and the portability, will be affected.

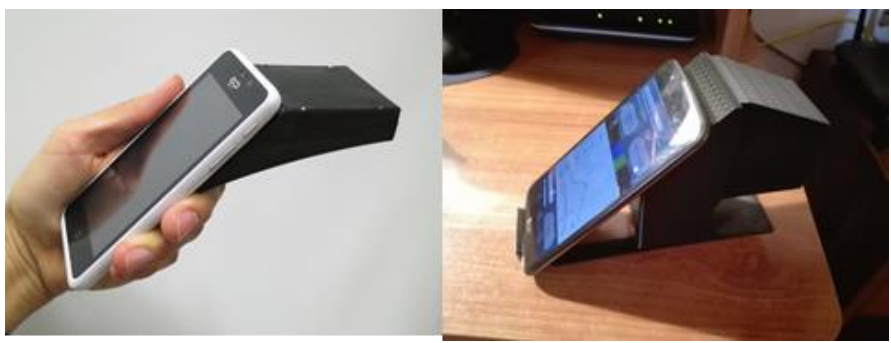

Fig. 7. Comparison of two peripheral. (a) WorkDevice Spectral Workbench device, and (b) our approach

Therefore, this solution, although interesting from the functionality of the spectrophotometer perspective, is not suitable when thinking about a solution for academic environments. On the one hand, a solution that forces students or teachers to sacrifice their phones to perform experiments is not viable. On the other hand, the alternative of buying smartphones solely and exclusively for the implementation of the spectrophotometer is neither appropriate, since it distorts the purpose of the project and unnecessarily rises the cost of the system.

In the literature, there are also other approaches, but they require from elements that are more sophisticated and do not take advantage of all the features of the smartphones. This is the case of the PASCO Wireless Spectrometer (https://www.pasco.com/spectrometer/), which creates a peripheral for the measurement of the data and sends them to the smartphone via bluetooth or USB connection. This type of solutions do not use any characteristic of the smartphone beyond handling and representation of data, since data is generated and obtained inside the device they have designed. Thus, the mobile is only a complement, and not a substitute for a scientific tool. Besides, these sophisticated peripherals unnecessarily increase the cost of the whole solution.

Regarding the cost efficiency, we have compared the cost of our proposal to the cost of the spectrophotometer used for the tests: Shimadzu UV-1800 [18], with an approximate market price of 16,750 USD (13,615 EUR). In order to estimate the cost of our approach, we take into account the cost of the PVC needed for printing the peripheral for a class with 20 students, plus the cost of the magnets and the diffraction gratings. All this material would approximately cost 59 EUR, which entails savings of more than 99%.

Of course, it would be argued that for a fair comparison, it is necessary to take also into account the cost of software development. However, our intention is providing this software under the license FOSS (Free and Open-Source Software), following the underlying philosophy of our software sources (Spectral Workbench and Apache Cordova). In fact, the software is currently available in GitHub, a web-based hosting service for version control using Git (https://github.com/CarlosUvigo/eSpectra).

Additionally, it should be borne in mind that with our proposal the costs of maintenance and repair of the scientific tool are also extremely reduced. Finally, increasing the number of users will reduce the cost of printing the case even more. Thus, our approach is undoubtedly a cost-effective solution compared to the use of the traditional laboratory spectrophotometer.

## 6. Conclusions and Future Work

Our approach to provide a VIS spectrophotometer for academic purposes by using a smartphone and an attachable peripheral is a complete and viable solution. It entails an important reduction in equipment costs and, additionally, do not requires having smartphones exclusively devoted for this aim. On the contrary, the removable peripheral allows obtaining the images with enough quality to be processed within the smartphone and to be represented with good quality, providing reliable results.

Although the design of the peripheral required from a detail analysis to be properly printed to maintain the 45º required to capture the images, the most complex part was undoubtedly adapting the software. For this to be possible, we needed to analyze the web-based system in deep to adapt the web-based system to the mobile phone without using its native language and, consequently, provide a horizontal solution ready to be used on any smartphone (independently from its software and/or hardware). It should be remarked that we used open source technologies, like Apache Cordova and its plugins and the code provided by Spectral Workbench.

Although our proposal has solved the majority of the challenges we faced at the beginning, we are currently working on improving some aspects. Firstly, we are trying to integrate the light source in the peripheral to provide a more stable source. Secondly, we are integrating some image processing features with the aim of (i) automatically manipulating the obtained images to decide the orientation that gives the best results and (ii) enhancing the image quality on the fly to increase the accuracy of the results.

**ACKNOWLEDGMENT**

This work is funded by: the European Regional Development Fund (ERDF) and the Galician Regional Government under agreement for funding the Atlantic Research Center for Information and Communication Technologies (AtlantTIC), and the Spanish Ministry of Economy and Competitiveness under the National Science Program (TEC2014-54335-C4-3-R and TEC2017-84197-C4-2-R).